\documentclass[12pt]{article}
\usepackage{axodraw}

\catcode`@=12
\begin{document}
\thispagestyle{empty}
\begin{flushright} 
UCRHEP-T363\\ 
August 2003\
\end{flushright}
\vspace{0.5in}
\begin{center}
{\LARGE	\bf Tripartite Neutrino Mass Matrix\\}
\vspace{1.5in}
{\bf Ernest Ma\\}
\vspace{0.2in}
{\sl Physics Department, University of California, Riverside, 
California 92521\\}
\vspace{1.5in}
\end{center}
\begin{abstract}\
The $3 \times 3$ Majorana neutrino mass matrix is written as a sum of 3 terms, 
i.e. ${\cal M}_\nu = {\cal M}_A + {\cal M}_B + {\cal M}_C$, where ${\cal M}_A$ 
is proportional to the identity matrix and ${\cal M}_{B,C}$ are invariant 
under different $Z_3$ transformations.  This ${\cal M}_\nu$ is very suitable 
for understanding atmospheric and solar neutrino oscillations, with 
$\sin^2 2 \theta_{atm}$ and $\tan^2 \theta_{sol}$ fixed at 1 and 0.5 
respectively, in excellent agreement with present data.  It has in fact 
been proposed before, but only as an ansatz.  This paper uncovers its 
underlying symmetry, thus allowing a complete theory of leptons and quarks 
to be constructed.

\end{abstract}
\newpage
\baselineskip 24pt

With the recent experimental progress in measuring atmospheric \cite{atm} 
and solar \cite{sol} neutrino oscillations, the mass-squared differences of 
the 3 active neutrinos and their mixing angles are now known with some 
precision.  Typical values at 90\% confidence level are \cite{valle}
\begin{eqnarray}
(\Delta m^2)_{atm} \sim (1.3 - 3.0) \times 10^{-3} ~{\rm eV}^2, &~& \sin^2 
2 \theta_{atm} \sim 0.88 - 1, \\ 
(\Delta m^2)_{sol} \sim (6 - 9) \times 10^{-5} ~{\rm eV}^2, &~& \tan^2 
\theta_{sol} \sim 0.33 - 0.76.
\end{eqnarray}
These few numbers have inspired the writing of hundreds of papers on the 
structure of the resulting $3 \times 3$ Majorana neutrino mass matrix 
${\cal M}_\nu$. Is the problem that complicated?  Perhaps not, if it is 
looked at with the proper perspective.

Motivated by the idea that ${\cal M}_\nu$ should satisfy \cite{ma03,mara03}
\begin{equation}
U {\cal M}_\nu U^T = {\cal M}_\nu,
\end{equation}
where $U$ is a specific unitary matrix, a very simple form of ${\cal M}_\nu$ 
is here proposed:
\begin{equation}
{\cal M}_\nu = {\cal M}_A + {\cal M}_B + {\cal M}_C,
\end{equation}
where
\begin{equation}
{\cal M}_A = A \pmatrix {1 & 0 & 0 \cr 0 & 1 & 0 \cr 0 & 0 & 1}, ~~ 
{\cal M}_B = B \pmatrix {-1 & 0 & 0 \cr 0 & 0 & -1 \cr 0 & -1 & 0}, 
~~ {\cal M}_C = C \pmatrix {1 & 1 & 1 \cr 1 & 1 & 1 \cr 1 & 1 & 1}.
\end{equation}
Since the invariance of ${\cal M}_A$ requires only $U_A U_A^T = 1$, $U_A$ can 
be any orthogonal matrix.  As for ${\cal M}_B$ and ${\cal M}_C$, they are 
both invariant under the $Z_2$ transformation \cite{allp,k03}
\begin{equation}
U_2 = \pmatrix {1 & 0 & 0 \cr 0 & 0 & 1 \cr 0 & 1 & 0}, ~~~ U_2^2 = 1,
\end{equation}
and each is invariant under a $Z_3$ transformation, i.e. 
$U_B^3 = 1$ and $U_C^3 = 1$, but $U_B \neq U_C$.  Specifically,
\begin{equation}
U_B = \pmatrix {-1/2 & -\sqrt{3/8} & -\sqrt{3/8} \cr \sqrt{3/8} & 1/4 & 
-3/4 \cr \sqrt{3/8} & -3/4 & 1/4}, ~~~  
U_C = \pmatrix {0 & 1 & 0 \cr 0 & 0 & 1 \cr 1 & 0 & 0}.
\end{equation}
Note that $U_B$ commutes with $U_2$, but $U_C$ does not. If $U_C$ is combined 
with $U_2$, then the non-Abelian discrete symmetry $S_3$ is generated.

First consider $C=0$.  Then ${\cal M}_\nu = {\cal M}_A + {\cal M}_B$ is the 
most general solution of
\begin{equation}
U_B {\cal M}_\nu U_B^T = {\cal M}_\nu,
\end{equation}
and the eigenvectors of ${\cal M}_\nu$ are $\nu_e$, $(\nu_\mu + \nu_\tau)/
\sqrt 2$, and $(\nu_\mu - \nu_\tau)/\sqrt 2$ with eigenvalues $A-B$, $A-B$, 
and $A+B$ respectively.  This explains atmospheric neutrino oscillations 
with $\sin^2 2 \theta_{atm} = 1$ and
\begin{equation}
(\Delta m^2)_{atm} = (A+B)^2 - (A-B)^2 = 4BA.
\end{equation}

Now consider $C \neq 0$.  Then in the basis spanned by $\nu_e$, $(\nu_\mu + 
\nu_\tau)/\sqrt 2$, and $(\nu_\mu - \nu_\tau)/\sqrt 2$,
\begin{equation}
{\cal M}_\nu = \pmatrix {A-B+C & \sqrt 2 C & 0 \cr \sqrt 2 C & A-B+2C & 0 \cr 
0 & 0 & A+B}.
\end{equation}
The eigenvectors and eigenvalues become
\begin{eqnarray}
\nu_1 = {1 \over \sqrt 6} (2 \nu_e - \nu_\mu - \nu_\tau), && m_1 = A-B, \\ 
\nu_2 = {1 \over \sqrt 3} (\nu_e + \nu_\mu + \nu_\tau), && m_2 = A-B+3C, \\ 
\nu_3 = {1 \over \sqrt 2} (\nu_\mu - \nu_\tau), && m_3 = A+B.
\end{eqnarray}
This explains solar neutrino oscillations as well with $\tan^2 \theta_{sol} = 
1/2$ and
\begin{equation}
(\Delta m^2)_{sol} = (A-B+3C)^2 - (A-B)^2 = 3C(2A-2B+3C).
\end{equation}
Whereas the mixing angles are fixed, the proposed ${\cal M}_\nu$ has the 
flexibility to accommodate the three patterns of neutrino masses often 
mentioned, i.e.

(I) the hierarchical solution, e.g. $B=A$ and $C << A$;

(II) the inverted hierarchical solution, e.g. $B = -A$ and $C << A$; 

(III) the degenerate solution, e.g. $C << B << A$.

\noindent In all cases, $C$ must be small. Therefore ${\cal M}_\nu$ of Eq.~(4) 
satisfies Eq.~(8) to a very good approximation, and $Z_2 \times Z_3$ as 
generated by $U_2$ and $U_B$ should be taken as the underlying symmetry of 
this model.

Since ${\cal M}_C$ is small and breaks the symmetry of ${\cal M}_A + 
{\cal M}_B$, it is natural to think of its origin in terms of the well-known 
dimension-five operator \cite{w79}
\begin{equation}
{\cal L}_{eff} = {f_{ij} \over 2 \Lambda} (\nu_i \phi^0 - l_i \phi^+)(\nu_j 
\phi^0 - l_j \phi^+) + H.c.,
\end{equation}
where $(\phi^+,\phi^0)$ is the usual Higgs doublet of the Standard Model and 
$\Lambda$ is a very high scale.  As $\phi^0$ picks up a nonzero vacuum 
expectation value $v$, neutrino masses are generated, and if $f_{ij} v^2/
\Lambda = C$ for all $i,j$, ${\cal M}_C$ is obtained.  Since $\Lambda$ is 
presumably of order $10^{16}$ to $10^{18}$ GeV, $C$ is of order $10^{-3}$ 
to $10^{-5}$ eV.  Using Eq.~(14) and Eq.~(2), $A-B+3C/2$ is then of order 
$10^{-2}$ to 1 eV.  This range of values is just right to encompass all 
three solutions mentioned above.

As for the form of ${\cal M}_C$, it may be understood as coming from effective 
universal interactions among the leptons at the scale $\Lambda$.  For example, 
if Eq.~(15) has $S_3$ symmetry as generated by $U_2$ and $U_C$, the most 
general form of ${\cal M}_C$ would be
\begin{equation}
{\cal M}_C = C \pmatrix {1 & 1 & 1 \cr 1 & 1 & 1 \cr 1 & 1 & 1} + 
C' \pmatrix {1 & 0 & 0 \cr 0 & 1 & 0 \cr 0 & 0 & 1}.
\end{equation}
However, the $C'$ term can be absorbed into ${\cal M}_A$, so again 
${\cal M}_\nu$ of Eq.~(4) is obtained.  This form of the neutrino mass 
matrix has in fact been discussed as an ansatz in a number of recent 
papers \cite{hps02,xing,hs02,hs03,hz}.

Consider now ${\cal M}_\nu$ of Eq.~(4) rewritten as
\begin{equation}
{\cal M}_\nu = (A+C) \pmatrix {1 & 0 & 0 \cr 0 & 1 & 0 \cr 0 & 0 & 1} - 
B \pmatrix {1 & 0 & 0 \cr 0 & 0 & 1 \cr 0 & 1 & 0} + C \pmatrix {0 & 1 & 0 \cr 
0 & 0 & 1 \cr 1 & 0 & 0} + C \pmatrix {0 & 0 & 1 \cr 1 & 0 & 0 \cr 0 & 1 & 0}.
\end{equation}
Note that each of the above four matrices is a group element of $S_3$.  This 
is the recent proposal of Harrison and Scott \cite{hs03}.  The difference 
here is that the underlying symmetry of ${\cal M}_\nu$ has been identified, 
thus allowing a complete theory of leptons and quarks to be constructed.

Going back to $U_B$ of Eq.~(7), its eigenvectors and eigenvalues are
\begin{eqnarray}
{1 \over \sqrt 2} (\nu_\mu - \nu_\tau), && \lambda_1 = 1, \\ 
{i \over \sqrt 2} \nu_e + {1 \over 2} (\nu_\mu + \nu_\tau), && \lambda 
= \omega, \\ 
{i \over \sqrt 2} \nu_e - {1 \over 2} (\nu_\mu + \nu_\tau), && \lambda 
= \omega^2,
\end{eqnarray}
where $\omega = e^{2 \pi i/3}$.  To accommodate this $Z_3$ symmetry in a 
complete theory, the Standard Model of particle interactions is now extended 
\cite{mara03} to include three scalar doublets $(\phi^0_i, \phi^-_i)$ 
and one very heavy triplet $(\xi^{++}, \xi^+, \xi^0)$.  The leptonic Yukawa 
Lagrangian is given by
\begin{eqnarray}
{\cal L}_Y = h_{ij} [\xi^0 \nu_i \nu_j - \xi^+ (\nu_i l_j + l_i \nu_j)/
\sqrt 2 + \xi^{++} l_i l_j] 
+ f_{ij}^k (l_i \phi^0_j - \nu_i \phi^-_j) l^c_k + H.c.,
\end{eqnarray}
where, under the $Z_3$ transformation,
\begin{eqnarray}
&& (\nu,l)_i \to (U_B)_{ij} (\nu,l)_j, ~~~ l^c_k \to l^c_k, \\ 
&& (\phi^0,\phi^-)_i \to (U_B)_{ij} (\phi^0,\phi^-)_j, ~~~ (\xi^{++}, \xi^+, 
\xi^0) \to (\xi^{++}, \xi^+, \xi^0).
\end{eqnarray}
This means
\begin{equation}
U_B^T h U_B = h, ~~~ U_B^T f^k U_B = f^k,
\end{equation}
resulting in
\begin{equation}
h = \pmatrix {a-b & 0 & 0 \cr 0 & a & -b \cr 0 & -b & a}, ~~~ f^k = \pmatrix 
{a_k - b_k & d_k & d_k \cr -d_k & a_k & -b_k \cr -d_k & -b_k & a_k}.
\end{equation}
Note that $h$ has no $d$ terms because it has to be symmetric.  Note also 
that both $h$ and $f$ are invariant under $U_2$ of Eq.~(6). Whereas the 
neutrino mass matrix ${\cal M}_A + {\cal M}_B$ is obtained with $A = 2a 
\langle \xi^0 \rangle$ and $B = 2b \langle \xi^0 \rangle$, the charged-lepton 
mass matrix ${\cal M}_l$ linking $l_i$ to $l^c_k$ has each of its 3 columns 
given by
\begin{equation}
({\cal M}_l)_{ik} = \pmatrix {(a_k-b_k)v_1 + d_k(v_2+v_3) \cr -d_k v_1 + 
a_k v_2 - b_k v_3 \cr -d_k v_1 - b_k v_2 + a_k v_3},
\end{equation}
where $v_i \equiv \langle \phi_i^0 \rangle$.  Assume $d_k,b_k << a_k$ and 
$v_{1,2} << v_3$, then all elements in the first, second, and third rows are 
of order $a_k v_1 + d_k v_3$, $a_k v_2 - b_k v_3$, and $a_k v_3$ 
respectively.  It is clear that they may be chosen to be of order $m_e$, 
$m_\mu$, and $m_\tau$, in which case ${\cal M}_l$ will become nearly 
diagonal by simply redefining the $l^c_k$ basis.  The mixing matrix $V_L$ 
in the $l_i$ basis (such that $V_L {\cal M}_L {\cal M}_L^\dagger V_L^\dagger$ 
is diagonal) will be very close to the identity matrix with off-diagonal 
terms of order $m_e/m_\mu$, $m_e/m_\tau$, and $m_\mu/m_\tau$.  This 
construction allows ${\cal M}_\nu$ of Eq.~(4) to be in the $(\nu_e, \nu_\mu, 
\nu_\tau)$ basis as a very good approximation.  The small deviation is also 
desirable for obtaining a nonzero but small value of $U_{e3}$, which is 
restricted by reactor data \cite{react} to be less than about 0.16 in 
magnitude. The consequences of having three Higgs doublets in this model 
are very similar to those discussed in Ref.~[5] and are repeated here below.

The Yukawa couplings of the three Higgs doublets are given by Eq.~(25). 
Taking the limit that only $v_3$ is nonzero, the charged-lepton mass matrix 
is simply given by
\begin{equation}
{\cal M}_l = v_3 \pmatrix {d_1 & d_2 & d_3 \cr -b_1 & -b_2 & -b_3 \cr 
a_1 & a_2 & a_3},
\end{equation}
whereas $\phi_1^0$ and $\phi_2^0$ couple to $l_i l_j^c$ according to
\begin{equation}
\pmatrix {a_1-b_1 & a_2-b_2 & a_3-b_3 \cr -d_1 & -d_2 & -d_3 \cr -d_1 & -d_2 
& -d_3}, ~~~ \pmatrix{d_1 & d_2 & d_3 \cr a_1 & a_2 & a_3 \cr -b_1 & -b_2 
& -b_3},
\end{equation}
respectively.  Assuming the hierarchy $d_k << b_k << a_k$ and rotating 
${\cal M}_l$ of Eq.~(27) in the $l^c_j$ basis to define the state 
corresponding to $\tau$, it is clear from Eq.~(28) that the dominant 
coupling of $\phi_1^0$ is $(m_\tau/v_3) e \tau^c$ and that of 
$\phi_2^0$ is $(m_\tau/v_3)\mu \tau^c$.  Other couplings are at most of 
order $m_\mu/v_3$ in this model, and some are only of order $m_e/v_3$.  
The smallness of flavor changing decays in the leptonic sector is thus 
guaranteed, even though they should be present and may be observable in 
the future.

Using Eq.~(28), we see that the decays $\tau^- \to e^- e^+ e^-$ and $\tau^- 
\to e^- e^+ \mu^-$ may proceed through $\phi_1^0$ exchange with coupling 
strengths of order $m_\mu m_\tau/v_3^2 \simeq (g^2/2) (m_\mu m_\tau/M_W^2)$, 
whereas the decays $\tau^- \to \mu^- \mu^+ \mu^-$ and $\tau^- \to \mu^- \mu^+ 
e^-$ may proceed through $\phi_2^0$ exchange also with coupling strengths of 
the same order.  We estimate the order of magnitude of these branching 
fractions to be
\begin{equation}
B \sim \left( {m_\mu^2 m_\tau^2 \over m_{1,2}^4} \right) B(\tau \to \mu \nu 
\nu) \simeq 6.1 \times 10^{-11} \left( {100~{\rm GeV} \over m_{1,2}} \right)^4,
\end{equation}
which is well below the present experimental upper bound of the order 
$10^{-6}$ for all such rare decays \cite{pdg}.

The decay $\mu^- \to e^- e^+ e^-$ may also proceed through $\phi_1^0$ with 
a coupling strength of order $m_\mu^2/v_3^2$.  Thus
\begin{equation}
B(\mu \to e e e) \sim {m_\mu^4 \over m_1^4} \simeq 1.2 \times 10^{-12} \left( 
{100~{\rm GeV} \over m_1} \right)^4,
\end{equation}
which is at the level of the present experimental upper bound of $1.0 \times 
10^{-12}$.  The decay $\mu \to e \gamma$ may also proceed through $\phi_2^0$ 
exchange (provided that $Re \phi_2^0$ and $Im \phi_2^0$ have different masses) 
with a coupling of order $m_\mu m_\tau/v_3^2$.  Its branching fraction is 
given by \cite{mara01}
\begin{equation}
B(\mu \to e \gamma) \sim {3 \alpha \over 8 \pi} {m_\tau^4 \over m_{eff}^4},
\end{equation}
where
\begin{equation}
{1 \over m_{eff}^2} = {1 \over m_{2R}^2} \left( \ln {m_{2R}^2 \over m_\tau^2} 
- {3 \over 2} \right) - {1 \over m_{2I}^2} \left( \ln {m_{2I}^2 \over 
m_\tau^2} - {3 \over 2} \right).
\end{equation}
Using the experimental upper bound \cite{meg} of $1.2 \times 10^{-11}$, we 
find $m_{eff} > 164$ GeV.

In the quark sector, if we use the same 3 Higgs doublets for the corresponding 
Yukawa couplings, the resulting $up$ and $down$ mass matrices will be of the 
same form as Eq.~(26).  Because the quark masses are hierarchical in each 
sector, we will also have nearly diagonal mixing matrices as in the case of 
the charged leptons.  This provides a qualitative understanding in our model 
of why the charged-current mixing matrix linking $up$ quarks to $down$ quarks 
has small off-diagonal entries.

Once $\phi_1^0$ or $\phi_2^0$ is produced, its dominant decay will be to 
$\tau^\pm e^\mp$ or $\tau^\pm \mu^\mp$ if each couples only to leptons. 
If they also couple to quarks (and are sufficiently heavy), then the dominant 
decay products will be $t \bar u$ or $t \bar c$ together with their 
conjugates.  As for $\phi_3^0$, it will behave very much as the single 
Higgs doublet of the Standard Model, with mostly diagonal couplings to 
fermions.  It should also be identified with the $\phi$ of Eq.~(15).

In conclusion, a simple tripartite neutrino mass matrix has been proposed, 
where the first two parts, ${\cal M}_A + {\cal M}_B$, are invariant under 
$U_B$ of Eq.~(7) with $U_B^3 = 1$.  The third part ${\cal M}_C$ is 
considered as a small perturbation which is democratic in the 
$(\nu_e,\nu_\mu,\nu_\tau)$ basis.  The resulting sum of the three parts is 
invariant under $U_2$ of Eq.~(6) with $U_2^2 = 1$ and fixes 
$\sin^2 2 \theta_{atm} = 1$ and $\tan^2 \theta_{sol} = 0.5$; but it also 
has 3 free parameters $A,B,C$ which determine the 3 neutrino mass eigenvalues 
as given in Eqs.~(11) to (13).  This structure with the underlying symmetry 
$Z_3 \times Z_2$ is supported in the context of a complete theory of leptons 
(that may be extended to quarks) which includes one very heavy Higgs triplet 
and three Higgs doublets at the electroweak scale, with experimentally 
verifiable properties.\\

This work was supported in part by the U.~S.~Department of Energy
under Grant No.~DE-FG03-94ER40837.

\bibliographystyle{unsrt}

\end{document}